\documentclass[twoside,12pt]{amsart}

\usepackage[foot]{amsaddr}
\usepackage{ulem}
\usepackage[frozencache,cachedir=.]{minted}
\usepackage{placeins}

\author{Simon Wilshin}
\address[Wilshin]{Royal Vet College, London, UK; ORCHID 0000-0002-8808-6659}

\author{Matthew D. Kvalheim}
\address[Kvalheim]{University of Pennsylvania, Philadelphia, PA, USA; ORCHID 0000-0002-2662-6760}

\author{Shai Revzen}
\address[Revzen, corresponding author]{University of Michigan, Ann Arbor, MI, USA; ORCHID 0000-0002-2989-0356}
\email{swilshin@rvc.ac.uk, kvalheim@seas.upenn.edu, shrevzen@umich.edu}

\title{Phase Response Curves and the Role of Coordinates}

\usepackage{import} 

\usepackage[utf8]{inputenc}
\usepackage[T1]{fontenc}
\usepackage{lmodern}
\usepackage{amsmath}
\usepackage{amssymb}
\usepackage{amsthm}
\usepackage{stmaryrd} 
\usepackage{mathtools}
\usepackage{tikz-cd}
\usepackage{subfig}

\usepackage{thmtools}
\usepackage{thm-restate} 

\newcommand{\concept}[1]{\textbf{#1}}

\newcommand{\R}{\mathbb{R}}
\newcommand{\C}{\mathbb{C}}
\newcommand{\pExt}{d}

\newcommand{\slot}{\,\cdot\,} 

\newcommand{\T}{\mathsf{T}}
\newcommand{\D}{\mathsf{D}}

\newcommand{\Del}{R}
\newcommand{\pvec}{Z}

\newcommand{\ip}[2]{\langle #1, #2 \rangle}

\newcommand{\paren}[1]{\left(#1\right)}
\newcommand{\anglebracket}[1]{\langle#1\rangle}

\newtheorem{Assump}{Assumption}
\newtheorem*{Goal}{Goal}

\newcommand{\thistheoremname}{}
\newtheorem*{genericthm}{\thistheoremname}
{\renewcommand{\thistheoremname}{Theorem~\ref{#1}$'$}%
	\begin{genericthm}}
	{\end{genericthm}}

\theoremstyle{definition}
\newtheorem{Def}{Definition}
\newtheorem*{Def*}{Definition}
\newtheorem{Rem}{Remark}

\newtheorem{Ex}{Example}

\usepackage{marvosym}
\usepackage[
hypertexnames,%
citecolor=blue,%
colorlinks=true,%
linkcolor=red%
]{hyperref}
\usepackage[all]{hypcap}

\begin{document}
	
	\maketitle
	
	\begin{abstract}
		The ``Phase Response Curve'' (PRC) is a common tool used to analyze phase resetting in the natural sciences.
		We make the observation that the PRC with respect to a coordinate $y\in\R$ actually depends on the full choice of coordinates $(x,y)$, $x\in\R^d$.
		We give a coordinate-free definition of the PRC making this observation obvious.
		We show how by controlling $y$, using delay coordinates of $y$, and postulating the dynamics of $x$ as a function of $x$ and $y$, we can sometimes reconstruct the PRC with respect to the $(x,y)$ coordinates.
		This suggests a means for obtaining the PRC of, e.g., a neuron via a voltage clamp. 
	\end{abstract}	
	\vspace{2em}
	\paragraph*{\small{keywords: phase response, dynamical systems, neural oscillator, Fitzhugh-Nagumo, co-ordinate dependence }}
	
	\paragraph*{\small{The authors have no relevant financial or non-financial interests to disclose.}}
	
	\tableofcontents
	\newcommand{\pDer}[1]{\partial_{#1}}
	\newcommand{\tDer}[1]{\frac{d}{d{#1}}}
	
	\newpage
	
	\section{Oscillators in the physical sciences}
	Many physical systems exhibit stable, long-term oscillations.
	Such oscillations are often modeled with ordinary differential equations that admit a periodic solution.
	The very definition of a periodic solution is such that, when following that solution, the state is entirely represented by a ``phase'' which is often taken to be an angle in radians or a fraction of a cycle. 
	
	A first-order model of how such system dynamics respond to external influences is given by the ``Phase Response Curve (PRC)'' which represents the ratio of infinitesimal phase change to infinitesimal external perturbation.
	
	It should be noted that the term ``PRC'' is also used in some fields, e.g. neuroscience, to describe the non-infinitesimal change in phase produced by applying a known limited-duration perturbation at various phases; we refer to this as a  non-infinitesimal PRC.
	The PRC of the previous paragraph is not itself sufficient to predict a non-infinitesimal PRC unless the perturbation is so small that the response is governed by its first-order approximation.
	The current paper focuses on the (infinitesimal) PRC of the previous paragraph, and we will use the term ``PRC'' for it from here on.
	
	\section{Coordinate-dependence of the standard PRC definition}\label{sec:standard}
	Consider the ordinary differential equation (ODE)
	$$\dot{x} = f(x)$$
	with $x(t) \in \R^n$, where $\dot{x}\coloneqq dx/dt$.
	Assume that $f$ is a $C^{1}$ vector field on $\R^n$ having a stable hyperbolic $T$-periodic orbit with image $\Gamma$ and basin of attraction $B\subset \R^n$.
	Then there exists \cite{shGuc75} a unique (modulo rotations) $C^1$ \concept{asymptotic phase map} $\varphi\colon B\to S^1\subset \C$ defined by the property that $$\dot\varphi = \frac{2\pi i}{T}\varphi$$
	where $i= \sqrt{-1}$.
	Equivalently, $\tDer{t}\arg(\varphi) = \frac{2\pi}{T}$ wherever $\arg(\varphi)$ is differentiable.
	
	If $(x_1,x_2,\ldots,x_n)$ are coordinates for $\R^n$, then the standard definition of the (infinitesimal) phase response curve (PRC) with respect to, say, $x_1$ is a map $\rho_1\colon \Gamma\to \R$ defined by the partial derivative $\pDer{x_1} := \frac{\partial}{\partial x_1}$ as 
	$$\rho_1\coloneqq \frac{1}{i\varphi}\pDer{x_1}\varphi.$$
	Equivalently, $\rho_1(x) = \pDer{x_1}\arg(\varphi(x))$ wherever $\arg(\varphi(\cdot))$ is differentiable.
	See, e.g., \cite{canavier2006} and the references therein.
	
	Suppose now that the last $(n-1)$ coordinates are modified to produce a coordinate system of the form $(x_1,y_2,\ldots,y_n)$.
	We can again define a PRC with respect to $x_1$, now denoted $\tilde{\rho}_1\colon \Gamma \to \R$, via
	$$\tilde{\rho}_1\coloneqq \frac{1}{i\varphi}\pDer{x_1}\varphi.$$
	Let us point out that, in general, $\rho_1\neq \tilde{\rho}_1$.
	In the next section we introduce a coordinate-free definition of the PRC which makes this observation obvious.
	As a preliminary explanation, we conclude this section with the following chastening of V. I. Arnold \cite[p.~258, foot. 81]{arnol1989mathematical}:
	\begin{quote}
		``It is important to note that the quantity $\frac{\partial u}{\partial x}$ on the $x,y$-plane depends not only on the function which is taken for $x$, but also on the function $y$: in new variables $(x,z)$ the value of $\frac{\partial u}{\partial x}$ will be different.
		One should write
		$$\left.\frac{\partial u}{\partial x}\right|_{y = \textnormal{const.}} \qquad \left.\frac{\partial u}{\partial x}\right|_{z = \textnormal{const.}}.\textnormal{''}$$	
	\end{quote}
	
	\section{Coordinate-free definition of the PRC}\label{sec:coord-inv}
	We now reformulate things more geometrically using standard definitions from smooth manifold theory \cite{lee2013smooth}.
	Let $M$ be a (finite-dimensional) smooth manifold and $f$ be a $C^{1}$ vector field on $M$ having a stable hyperbolic $T$-periodic orbit with image $\Gamma$ and basin of attraction $B\subset M$.
	Then there exists \cite{shGuc75} a unique (modulo rotations) $C^1$ \concept{asymptotic phase map} $\varphi\colon B\to S^1\subset \C$ defined by the property that
	$$\ip{\varphi^*dx}{f} \equiv \frac{2\pi}{T},$$
	where $dx$ is the standard volume form on $S^1$ satisfying $\int_{S^1} dx = 1$, $\varphi^*dx$ is the pullback of $dx$, and $\ip{\slot}{\slot}$ denotes the $\R$-valued bilinear pairing between cotangent and tangent vectors.
	\begin{Def}\label{def:PRC}
		Let $\pvec\colon \Gamma\to \T M|_\Gamma$ assign to each $x\in \Gamma$ a vector $\pvec(x)\in \T_x M$.
		Then the \concept{$\pvec$-phase response curve (PRC)} $\rho_\pvec\colon \Gamma\to \R$ is defined by
		\begin{equation}
		\rho_\pvec(x)\coloneqq \ip{\varphi^*dx}{\pvec}(x).
		\end{equation}
	\end{Def}
	\begin{Rem}\label{rem:prc-defs-relationship}
		The connection of Def.~\ref{def:PRC} with the PRC definition of \S\ref{sec:standard} is as follows: take $M = \R^n$ and let $\pvec = \pDer{x_1}$ be the first coordinate vector field defined by a choice of coordinates $(x_1,x_2,\ldots,x_n)$ for $\R^n$.
		
		If the last $(n-1)$ coordinates are changed to produce a coordinate system $(x_1,y_2,\ldots, y_n)$, then with respect to these new coordinates the first coordinate vector field $\tilde{\pvec}$ is generally not equal to $\pvec$; it is the pushforward of $\pvec$ along the coordinate change $\Psi\colon (x_1,x_2,\ldots, x_n)\mapsto (x_1,y_2,\ldots, y_n)$.
		Explicitly, $$\tilde{\pvec} = \Psi_* \pvec\coloneqq \D \Psi \circ \pvec \circ \Psi^{-1}.$$
		In light of Def.~\ref{def:PRC} and Rem.~\ref{rem:prc-defs-relationship}, this explains the observation made at the end of \S\ref{sec:standard}.
	\end{Rem}
	\begin{Rem}\label{rem:coord-change}
		More generally, let $\Psi\colon M\to \tilde{M}$ be a diffeomorphism.
		The vector field $f$ pushes forward to a vector field $\tilde{f}= \Psi_* f$ on $\tilde{M}$ having a stable hyperbolic $T$-periodic orbit with image $\tilde{\Gamma}= \Psi(\Gamma)$, basin of attraction $\tilde{B} = \Psi(B)$, and asymptotic phase map $\tilde{\varphi}\colon \tilde{B}\to S^1$ given by $\tilde{\varphi} = \varphi\circ \Psi^{-1}$. 
		Given $\pvec\colon \Gamma\to \T M|_\Gamma$, we define $\tilde{\pvec}\colon \tilde{\Gamma}\to \T \tilde{M}|_{\tilde{\Gamma}}$ by $\tilde{\pvec}\coloneqq \D \Psi \circ \pvec \circ \Psi^{-1}|_{\tilde{\Gamma}}$.
		The definitions of pullbacks and pushforwards immediately imply that, on $\tilde{\Gamma}$:
		\begin{equation}
		\rho_\pvec \circ \Psi^{-1} = \ip{\varphi^*dx}{\pvec}\circ \Psi^{-1} = \ip{(\Psi^{-1})^*\varphi^* dx}{\Psi_* \pvec} = \ip{(\varphi\circ \Psi^{-1})^*dx}{\Psi_*\pvec} = \ip{\tilde{\varphi}^*dx}{\tilde{\pvec}}\eqqcolon \tilde{\rho}_{\tilde{\pvec}}.
		\end{equation}
		Hence one can obtain the PRC $\rho_\pvec$ from the data on $\tilde{M}$ using $\rho_\pvec(x) = \tilde{\rho}_{\tilde{\pvec}}(\Psi(x))$.    
	\end{Rem}
	
	\begin{Ex}\label{ex:delay-1}
		Let $\Phi$ be the local flow generated by $f$; since we are only interested in dynamics on $B$, we may assume without loss of generality that $\Phi\colon \R\times M\to M$ is a flow.
		Given $0 = \delta_0 < \delta_1 < \delta_2 < \cdots < \delta_m$ and a smooth function $h\colon M\to \R$, define $\Del\colon M\to \R^{m+1}$ via
		\begin{equation}
		\Del(x)\coloneqq (h(x), h\circ \Phi^{-\delta_1}(x), h\circ \Phi^{- \delta_2}(x),\ldots, h\circ \Phi^{-\delta_m}(x)).
		\end{equation}
		Let us assume that $\Del$ is a smooth embedding so that the codomain-restricted map $\Psi\coloneqq \Del\colon M\to \Del(M)\eqqcolon \tilde{M}$ is a diffeomorphism.
		
		We can express PRCs for $f$ in terms of this data as follows using Rem.~\ref{rem:coord-change}: 
		given any vector field $\pvec\colon \Gamma \to \T M|_\Gamma$ over $\Gamma$, the $\pvec$-PRC $\rho_\pvec\colon \Gamma\to \R$ is given by
		\begin{equation}\label{eq:prc-delay-expression}
		\rho_\pvec(x) = \ip{\tilde{\varphi}^*dx}{\Psi_* \pvec}(\Psi(x)) = \ip{\tilde{\varphi}^*dx}{\sum_{k=0}^{m}(\D h\circ \D\Phi^{-\delta_k}\pvec(x)) \pDer{x_k}},
		\end{equation}
		where $\pDer{x_k}$ is the $k$-th coordinate vector field defined by the coordinates $(x_0,x_1,\ldots,x_{m})$ for $\R^{m+1}$.
	\end{Ex}
	
	\section{An extended example}\label{sec:ex-hopf}
	In this section we illustrate the observations of \S \ref{sec:standard} and \S \ref{sec:coord-inv} by instantiating Example~\ref{ex:delay-1} in a concrete setting.
	Consider a Hopf-like oscillator in polar coordinates on $\R^2\setminus \{\mathbf{0}\}$:
	\begin{align}
	\dot{\theta}=w &~& \dot{r}=1-r
	\end{align}
	where dots indicate differentiation with respect to time $t$. 
	This has the general solution:
	\begin{align}
	\theta=w t +\theta_0&~& r=1+\rho_0\exp(-t),
	\end{align}
	where $\rho_0 := r(0)-1$.
	Note that $\{r = 1\}$ is the image of an exponentially stable periodic orbit for this system; we will also refer to this periodic orbit as a ``limit cycle''.
	
	Let us switch to Cartesian coordinates and imagine that our $y$ coordinate is some variable we can observe (for example, a voltage of a neuron), while the $x$ coordinate is some variable we do not observe (for example, the associated current).
	We rewrite the solutions as:
	\begin{eqnarray}
	x & = & [1+\rho_0\exp(-t)]\cos\paren{w t+\theta_0} \\
	y & = & [1+\rho_0\exp(-t)]\sin\paren{w t+\theta_0}.
	\end{eqnarray}
	Recalling that $r = \sqrt{x^2+y^2}$, and noting that the phase of this oscillator can be taken to be $\theta$, we can compute the phase $1$-form in this coordinate system as:
	\newcommand{\dExt}{d}
	\begin{equation}\label{eq:dtheta}
	\dExt\theta(x,y) = \paren{-y/r^2,x/r^2},
	\end{equation}
	where the right side should be viewed as a row vector.\footnote{Despite the fact that the closed $1$-form defined by the right side of \eqref{eq:dtheta} is not exact on $\R^2\setminus \{\mathbf{0}\}$,  ``$d\theta$'' is still common notation for this $1$-form. See for example \cite[p.~93]{spivak1971calculus}.}
	Using for simplicity the terminology of \S \ref{sec:standard}, the restrictions to the unit circle of the components of this 1-form are the PRCs with respect to $x$ and $y$ in the coordinate system $(x,y)$. 
	We also know the isochrons explicitly, since the asymptotic phase map is given by $\theta$ and its level sets are radial lines.
	This gives rise to the coordinate-independent observation that a tangent vector $n$ is tangent to some isochron if, and only if, it satisfies $\anglebracket{\dExt\theta,n}=0$.
	In Cartesian coordinates this is satisfied for $n=\paren{n_0,n_1}$ if $y n_0 = x n_1$,	that is, if the vector is radial.
	
	As in Example~\ref{ex:delay-1}, we will now consider delay coordinates $(v,u)$ and how we obtain them from $(x,y)$. 
	The coordinate $u(t)$ is the same coordinate which we observe, $y(t)$.
	For the coordinate $v(t)$ we take the value of the $y$ coordinate of our system at some lag time $\delta>0$, i.e., $y(t-\delta)$.
	This gives
	\begin{align}
	u(t) := & ~y(t) = [1+\rho_0\exp(-t)]\sin(w t+\theta_0) \\
	v(t) := & ~y(t-\delta) = [1+\rho_0\exp(-t+\delta)]\sin(w t-w \delta+\theta_0)
	\end{align}
	We would like to know how the PRCs with respect to $v$ and $u$ in the coordinates $(v,u)$ are related to those we previously calculated with respect to the $(x,y)$ coordinates.
	In the present example we have an explicit formula for the coordinate transformation, so we can eliminate time from the above relations and carry out the remaining computations explicitly.
	The $u$ coordinate change is trivial since, by construction, $u=y$.
	
	Note the following identities and definitions, all written in terms of $x$, $y$, and $r:=\sqrt{x^2+y^2}$: $\exp(-t) = (r-1)/\rho_0$, $y/r = \sin(w t+\theta_0)$, $x/r = \cos(w t +\theta_0)$, 
	$C := \cos(w\delta)$, and  $S := \sin(w\delta)$.
	We thus have:
	\begin{align}
	    v &= \left[1+\rho_0 e^\delta \cdot (r-1)/\rho_0\right]\sin(w t -w \delta+\theta_0) \nonumber\\
	    &= \left[1+(r-1)e^\delta\right](\sin(w t +\theta_0) C -\cos(w t +\theta_0) S) \nonumber\\
	    &= \left[(1-e^\delta)+r e^\delta\right] \frac{1}{r}(y C - x S) \nonumber\\
	    &= \left[(1-e^\delta)/r+e^\delta\right](y C - x S)\label{eq:v-simplified}
	\end{align}

	\newcommand{\pxy}[3]{\ensuremath{\left.\frac{\partial {#1}}{\partial {#2}}\right|_{#3}}}
	From this transformation we can compute the Jacobian $J$ of the transformation $T_\delta:(x,y)\mapsto (v,u)$
	The components of the Jacobian are (using the notation ``$|_x$'' as shorthand for ``$|_{x= \textnormal{const.}}$'' and similarly for ``$|_y$''):
	\begin{eqnarray}
	\pxy{u}{x}{y} ~=~ 0; &&
	\pxy{v}{x}{y} ~=~
	\frac{1}{r^{3}} \left(- S r^{2} \left(e^\delta r - e^\delta + 1\right) + x \left(e^\delta - 1\right) \left(C y - S x\right)\right); \nonumber\\
	\pxy{u}{y}{x} ~=~ 1; && 
	\pxy{v}{y}{x} ~=~
	\frac{1}{r^{3}} \left(C r^{2} \left(e^\delta r - e^\delta + 1\right) + y \left(e^\delta - 1\right) \left(C y - S x\right)\right).
	\nonumber
	\end{eqnarray}
	For certain values of the parameters $w$ and $\delta$, the quantity $\pxy{v}{x}{y}$ is nowhere vanishing on the image of the limit cycle.\footnote{Setting $r = 1$ in $\pxy{v}{x}{y}$ and viewing the resulting expression as a quadratic function of $x$ with parameter $y$, the discriminant $\Delta$ of this quadratic is $\Delta = (e^\delta-1)^2C^2y^2-4(e^\delta-1)S^2$. Hence $\Delta < 0$ with $\delta > 0$ if and only if $(e^\delta-1)C^2y^2<4S^2$. This implies that, with $\delta > 0$, $\Delta < 0$ for all $(x,y)\in \{r=1\}$ if and only if $(e^\delta -1)C^2<4S^2.$ Since a quadratic has a real root if and only if its discriminant is nonnegative, the latter inequality is necessary and sufficient for the quantity $\pxy{v}{x}{y}$ to be nowhere vanishing on $\{r=1\}$. If it so happens that $w\delta\not \in \frac{\pi}{2}+\pi\mathbb{Z}$, then this necessary and sufficient condition reads $$e^\delta -1<  4\tan^2(w\delta).$$ Since $\tan^2(w \delta) = w^2 \delta^2 + o(\delta^5)$ as $\delta \to 0$ while $e^\delta - 1 = \delta + o(\delta)$, we see that, for this condition to hold, $w$ must be very large if $\delta$ is very small.} 
	For such parameter values, this leads to the invertibility of the Jacobian $J$ at each point on the limit cycle, with expressions for $J$ and $J^{-1}$ as follows:
	\begin{align}
	J(x,y)= \paren{\begin{array}{cc}
		\pxy{v}{x}{y} & \pxy{v}{y}{x} \\
		0 & 1 
		\end{array}},
		&&
	J^{-1}(x,y) = \paren{\begin{array}{cc}
		\left(\pxy{v}{x}{y}\right)^{-1} 
		& -\left(\pxy{v}{y}{x}\right) \left(\pxy{v}{x}{y}\right)^{-1} \\
		0 & 1 \\
		\end{array}}.
	\end{align}
	If additionally $\delta \not \in \frac{2\pi}{w} \mathbb{Z}$ so that  the restriction of $T_\delta$ to $\{r=1\}$ is invertible, it can be shown that the nonlinear map $T_\delta$ is smoothly invertible on some open neighborhood of $\{r=1\}$ \cite[p.~19, Ex.~10]{guillemin1974differential}. 
	Thus, for such parameter values, we may use the transformation law for 1-forms to transform $\pExt\theta(x,y)$ on a neighborhood of $\{r=1\}$ according to 
	\begin{align}
	    \pExt\theta_\delta(v,u) := \pExt\theta(T_\delta^{-1}(v,u)) J^{-1}(T_\delta^{-1}(v,u))
	\end{align}
	Since $u=y$, one might assume mistakenly that the change in phase with respect to an infinitesimal perturbation in $y$ in the $(x,y)$ coordinates is the same as the change in phase with respect to an infinitesimal perturbation in $u$ in the $(u,v)$ coordinates.
	However, the former is $x/r^2$, whereas the latter is $x/r^2 + y/r^2 \pxy{v}{y}{x} \left(\pxy{v}{x}{y}\right)^{-1}$.

\begin{figure}[ht]
    \includegraphics[width=0.9\textwidth]{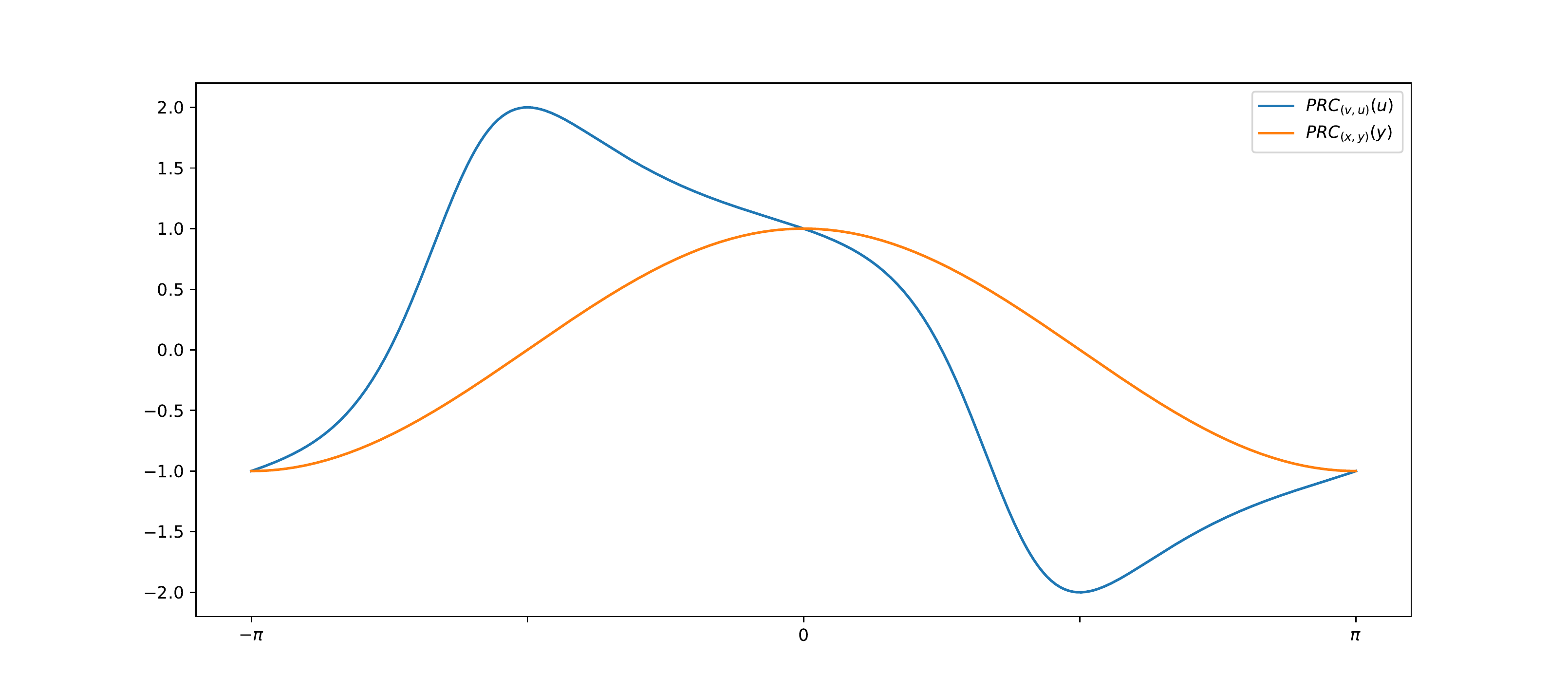}
    \caption{Comparison of the phase response with respect to an infinitesimal perturbation of the $u$-coordinate in $(v,u)$, and the $y$-coordinate in $(x,y)$. %
    Note that we have chosen $u=y$, and we are perturbing the same oscillator, with parameters $\delta := \ln 2$ and $w = \pi / (4 \ln 2)$. 
    }
    \label{fig:prc-example}
    \end{figure}

	Taking one specific example, choose $\delta := \ln 2$ and $w = \pi / (4 \ln 2)$, giving $C=S=1/\sqrt{2}$, and examine $J$ on the limit cycle $\{r=1\}$, setting $x:=\cos(\phi)$ and $y:=\sin(\phi)$.
	After some algebra, we obtain:
	
	\begin{align}
	  \pxy{v}{x}{y} & =-\frac{1}{2} \cos{\left (2 \phi + \frac{\pi}{4} \right )} - \frac{3 \sqrt{2}}{4}, \\
	  \pxy{v}{y}{x} & =-\frac{1}{2} \sin{\left (2 \phi + \frac{\pi}{4} \right )} + \frac{3 \sqrt{2}}{4}.
	\end{align}
	The resulting difference in PRCs between $y$ and $u$ is shown in Figure~\ref{fig:prc-example}.

	\section{PRC recovery with delay coordinates}\label{sec:prc-recovery-procedure}
	Eq. \eqref{eq:prc-delay-expression} suggests that the prospect of computing a desired PRC from measurements of a scalar-valued observable is grim, since \eqref{eq:prc-delay-expression} involves the (usually unknown) derivatives $\D\Phi^{-\delta_k}$ of the flow.
	However, in this section we show that certain PRCs \emph{can} be recovered by measurements of a scalar-valued observable under certain additional assumptions.
	We are motivated by the problem of recovering a PRC for an experimental system which provides limited ability to both observe and control its state.
	
	\subsection{Assumptions and goal}
	
	We continue to make the same assumptions as in \S \ref{sec:coord-inv}, but we now further assume that $M=X\times Y$ is the product of two smooth manifolds.
	We write the vector field $f$ on $M$ as $f = (f_x,f_y)$ so that an integral curve $(x(t),y(t))$ of $f$ satisfies the ODE  $\dot x = f_x(x,y); \dot y = f_y(x,y)$.
	As is traditional in control theory, we think of $x$ as being a ``hidden state'' whereas $y$ is ``directly observable'', i.e., available for us to measure at any instant. 
	We let $\gamma\colon \R\to M$ be a specific solution of the ODE with image $\gamma(\R)=\Gamma$, the image of the assumed hyperbolic $T$-periodic orbit, so that $\gamma(T)=\gamma(0)$ and $\gamma|_{[0,T)}$ is injective.
	We use the notation $\gamma_x(t)\in X$, $\gamma_y(t)\in Y$ for the components of $\gamma$ so that $\gamma(t) = (\gamma_x(t),\gamma_y(t))$.
	We now record our goal and remaining assumptions explicitly.
	
	\begin{Assump}
		$\dim(Y)=1$. (We believe this assumption could be relaxed by defining ``vector-valued PRCs''.)
	\end{Assump}
	
	\begin{Assump}\label{assump:known-fx}
		$f_x(\slot,\slot)$ is known.	
	\end{Assump}
	
	\begin{Goal}
		Find a method to recover the $\pDer{y}$-PRC from $y$ measurements.
	\end{Goal}
	
	\begin{Rem}
		We will not assume knowledge of $f_y(\slot,\slot)$. 
	\end{Rem}

	\begin{Assump}\label{assump:delay-emb}
		There exist known $0< \delta_1 < \delta_2 < \cdots < \delta_m$ such that the delay map $\Del\colon M\to \R^m$ given by $\Del\colon (x(t),y(t))\mapsto (y(t),y(t-\delta_1),\ldots,y(t-\delta_m))\in \R^m$ is an embedding for all trajectories $(x(t),y(t))$.
	\end{Assump}
	\begin{Rem}
		Takens' theorem \cite{takens1981detecting,sauer1991embedology} tells us that such a collection $\{ \delta_i \}_{i=0}^m$ of delays can usually be easily found. 
		In fact, given some technical conditions are satisfied, almost all collections of delays will do.
	\end{Rem}

	\begin{Assump}\label{assump:control-auth}
	For any $C^1$ curve $u\colon [t_0,t_1]\to Y$  satisfying $u(t_0) = y(t_0)$, we can enforce $y(t) =u(t)$ for all $t\in [t_0,t_1]$.
	\end{Assump}
	\newcommand{\Xu}{$\mathbf{X}(u)$}
	\newcommand{\Xus}{$\mathbf{X}(u^*)$}
	\newcommand{\Xy}{$\mathbf{X}(y)$}%
	\newcommand{\Xg}{$\mathbf{X}(\gamma_y)$}%
	\newcommand{\Xd}{$\Delta(\gamma)$}%
	
	We refer to the following control system $$\dot{x} = f_x(x(t),u(t))$$
	as the \Xu\ system.
	The \Xu\ system is a control system with control $u$ if $u$ is unspecified, and is a nonautonomous differential equation if a specific $u(t)$ (such as $\gamma_y(t)$) is specified.
	
	\begin{Assump}\label{assump:reset}
		$\gamma_x(t)$ is globally asymptotically stable for the  \Xg\ system $\dot{x} = f_x(x(t),\gamma_y(t))$.
	\end{Assump}

	\begin{Rem}
		While not always true, it is commonly believed that many periodically forced systems exhibit stable periodic responses, at least in the context of certain classes of dissipative mechanical systems \cite{breunung2019does}.
		The nonautonomous \Xg\ system indeed has $\gamma_x(t)$ as a solution, so the question is whether the global (or even local) asymptotic stability in Assumption \ref{assump:reset} holds for a given system.
		The following example shows that the class of systems for which Assumption \ref{assump:reset} holds is at least not empty.
	\end{Rem}
	
	\begin{Ex}
	Consider the dynamics
	\begin{align*}
	\dot{x} &= -x\\
	\dot{y} &=1
	\end{align*}
	with $x\in X = \R$ and $y\in Y= S^1$ (the circle), so that $f_x(x,y)=-x$, $f_y(x,y)=1$, and $\gamma_x(t)\equiv 0$.
	There is a globally asymptotically stable limit cycle with image $\{0\}\times S^1$.
	Since $f_x$ does not depend on $y$, $\gamma_x$ is still globally asymptotically stable for the \Xg\ system, so Assumption~\ref{assump:reset} is satisfied.
	\end{Ex}
	
	\begin{Rem}\label{rem:samp}
        In general, we do not know the delay map $\Del$.
        However, any trajectory of \Xu\ that we compute with a segment of time $[s,t]$, $t-s \geq \delta_m$,  that evolves autonomously (i.e., such that we do not externally force the system during $[s,t]$) gives us a delay coordinate point by observing $y|_{[s,t]}$.
        If, furthermore, we know $x(t)$ then we have sampled the value of $\Del$ at the point $(x(t),y(t))$ and found it to be $(y(t),y(t-\delta_1),\ldots,y(t-\delta_m))$.
	\end{Rem}

	\begin{Assump}\label{assump:controllability}
		There exists a neighborhood $U\supset \Gamma$ and a neighborhood $V\ni \gamma(0)$ such that, for every $(x_f,y_f)\in U$ and $(x_0,y_0)\in V$, there is a known $C^1$ control $u\colon [0,N]\to Y$ satisfying $u(0) = y_0$, $u(N)= y_f$, and steering $x(0)= x_0$ to $x_f$ for the \Xu\ dynamics.
	\end{Assump} 
	\subsection{PRC recovery procedure}\label{sec:prc-recovery-procedure-2}
    Under these assumptions we propose to recover the PRC from delay coordinate measurements and $Y$ clamping experiments as follows.
    
    \subsubsection{Preparation}\label{sec:compute-gamma-y}
		Compute from $y$-observations over a long enough time an estimate of the period $T$ and $\gamma_y$. 
		This also provides us with the delay representation of $\Gamma$, $\Del(\Gamma)$, and the phase on the limit cycle itself, $\varphi|_\Gamma$.
	
	\subsubsection{Sampling}\label{sec:samp}
	Now consider a fixed $u$ and a time $N$.
	\begin{enumerate}	
	
	\item\label{step:reset}
	    Let the system run autonomously, producing $z(t) \coloneqq (x(t),y(t))$ for our (as of yet unknown) $x(t)$ and our directly observable $y(t)$.
	    After $\delta_m$ units of time, this makes $\Del(z(t))$ known by Assumption~\ref{assump:delay-emb}.
	    Continue autonomously until $\Del(z) = \Del(\gamma(0))$ to within desired accuracy.
	    The system has ``reset''; we therefore define this time as $t=0$.
	    
	\item\label{step:apply-y-equals-uz}
		Use Assumptions~\ref{assump:control-auth} and \ref{assump:controllability} to force $y(t) := u(t)$ for $t\in [0,N]$ so that $z(N) \in U \setminus \Gamma$.
		We will later select these $z(N)$ to comprise a collection of points that enable us to estimate partial derivatives of $\varphi(\cdot)$. 
	
	\item Let the system run autonomously for at least $\delta_m$ more time units.
	    $R(z(t))$ is known again for any $t>N+\delta_m$.
	    Thus we can determine when $z(t)$ returns to $\Gamma$ to within sufficiently good accuracy by checking $R(z)$ for membership in $R(\Gamma)$.
	    Assume the time at which this happens is $t_F$, i.e., we approximately have $z(t_F) \in \Gamma$.
	    We can approximately obtain $\varphi_F \coloneqq \varphi(z(t_F))$ because in \ref{sec:compute-gamma-y} we obtained the phase on $\Gamma$ from $R(z(t_F))$.
	    From this and the period we get the phases for the entire time segment as $\varphi(z(t)) = \exp(-i 2 \pi (t_F-t)/T) \varphi_F$.
    \end{enumerate}
    
    Consider now from these trajectories the subset of points $\Del(z(t))$ sufficiently close to $\Del(\Gamma)$ for which which we also have $\varphi(z(t))$.
    This set of points will allow us to estimate the $\pDer{y}$-PRC with respect to delay coordinates (cf. Arnold's chastening in \S\ref{sec:standard}). 
    
    Importantly, note that we have not used any knowledge about a specific $x$ or $f_x$ other than what is mentioned in Assumptions~\ref{assump:delay-emb}, \ref{assump:control-auth}, and \ref{assump:controllability}.
    However, at this time our data collection is complete.
    
    \subsubsection{Reconstructing $\Del$}
    
    Now choose a specific $f_x$ that would meet our assumptions.
	Using Assumptions~\ref{assump:known-fx} and \ref{assump:reset} and knowledge of $\gamma_y$ from \ref{sec:compute-gamma-y}, numerically integrate the \Xg\ system for a large time to determine  $\gamma_x(0)$ as accurately as desired.

    Consider again the process of \ref{sec:samp}.
    For each trajectory computed therein, starting with the initial condition $z(0)=\gamma(0)$ which is now fully known, integrate the control system \Xy\ to give the trajectory $z(t)$ explicitly.
	For the times $t>N+\delta_m$, the (now) known $x(t),\Del(z(t))$ provide samples of the map $\Del(\cdot, \cdot)$.
	Using this collection of samples, interpolate the map $\Del$ and its derivative. 
	This allows us to use \eqref{eq:prc-delay-expression} to recover the desired PRC in the $(x,y)$ coordinates.
	Note also that by reconstructing $\Del$ we have acquired the means to convert any geometric object between delay coordinates and the original coordinates.

	\section{An example from neuroscience}\label{sec:ex-fhn}
    Whereas the mathematical comments and examples in the former sections are perhaps noteworthy, they are hardly surprising to the mathematician.
    However, they could have significant practical applications.
    We therefore switch our focus to numerical approaches that could be applied to (possibly noisy) experimental data.
    For this reason, in this section we will analyze data produced from numerically simulating a model system as if we were experimentalists without knowledge of the underlying equations of motion.
    We will conduct this analysis based on plausible assumptions informed by numerical tests, highlighting the places where analytical or computer-assisted proofs would perhaps be nice to have, but are beyond the scope of typical experimental work. 

	Let us consider the Fitzhugh-Nagumo (FHN) system, often used as a model of a simple neuronal oscillator.
	A standard form of the equations for this system is:
	\begin{align}\label{eq:fhn}
	\dot{v} = c\paren{v + w - \frac{v^3}{3}+z} &~& \dot{w} = \frac{a - v - bw}{c}
	\end{align}
	Here voltage $v$, current $w$, and time have all been suitably normalised. 
	Differentiation with respect to time is indicated by a dot.
	We adopted the values $0.7$, $0.8$ and $3$ for $a$, $b$ and $\tau$ respectively.
	
    Time series data generated by numerically integrating \eqref{eq:fhn} at these parameter values strongly suggests that there is an exponentially stable limit cycle with a large basin of attraction; we henceforth assume this is the case without a proof.
    We modeled the limit cycle with two univariate splines (one for each coordinate; using the scipy command \texttt{UnivariateSpline}) based on $2000$ points with the number of knots set to interpolate through all points.
    
\begin{figure}[ht]
    \includegraphics[width=0.9\textwidth]{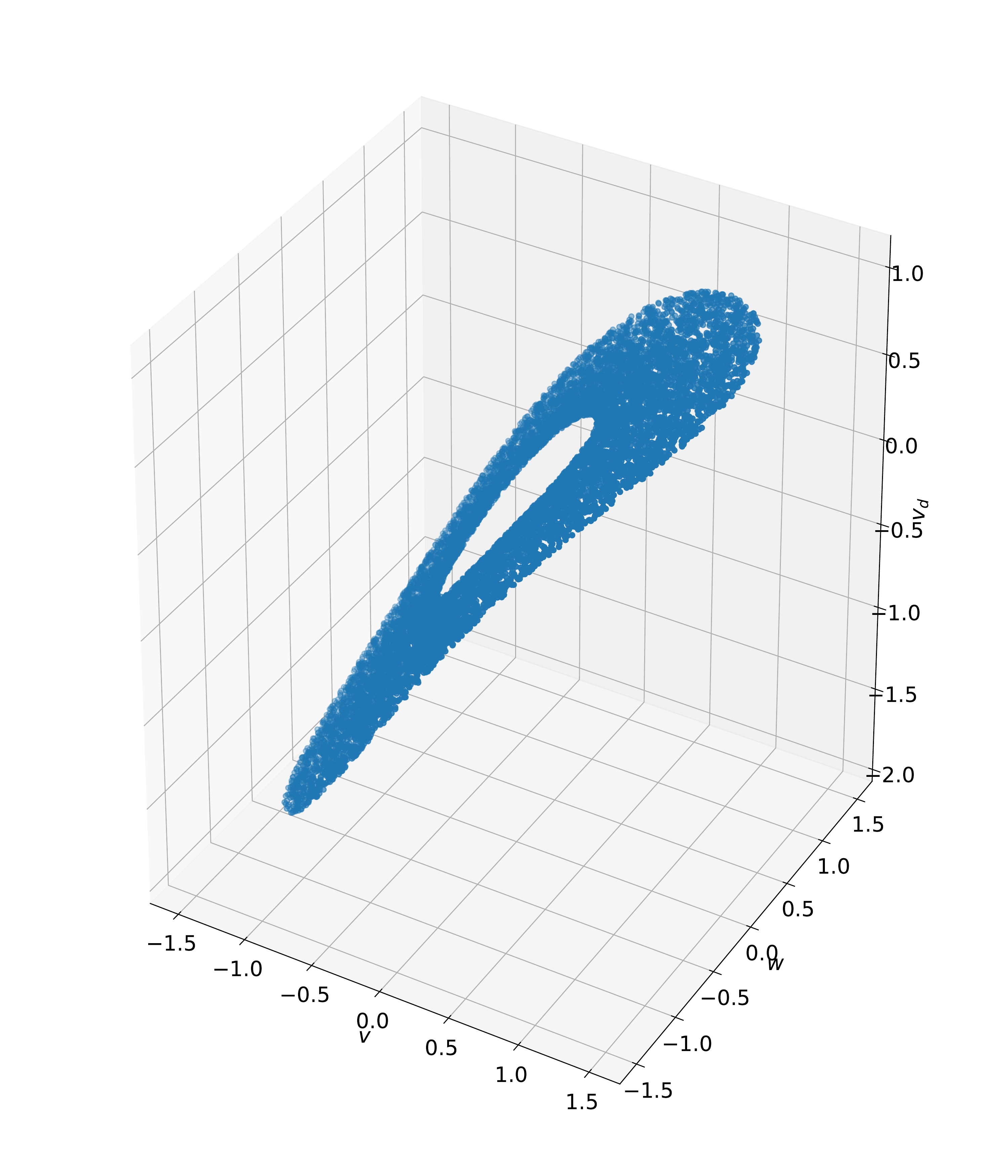}
    \caption{Shown here are sample points illustrating the coordinate transformation $(v,w)\mapsto (v,v_d)$ for the Fitzhugh-Nagumo system \eqref{eq:fhn} implicitly defined by delaying the $v$ coordinate. %
    Here $v$ is the normalised voltage, $w$ is the normalised current, and $v_d$ is the delayed normalised voltage.}
    \label{fig:delay-implicit-transform-just-v}
    \end{figure}

	Following the method of \S \ref{sec:prc-recovery-procedure}, we attempted to use delay coordinates $\paren{v,v_d}$ with a single delay of length $0.1$ seconds to build a full picture of the dynamics using only observations of $v$.
	The rationale behind using such a voltage-centered model is that, when observing a neuron, we are typically able to record its voltage but not the current flowing through it.
	The delay coordinates $\paren{v,v_d}$ are implicitly defined as functions of $(v,w)$:
	\begin{align}\label{eq:fhn-delay-coord-change}
	v = v & & v_d = g\paren{v,w}.
	\end{align}
	Using trajectory data, we generated an approximation of the graph of the map $(v,w)\mapsto g(v,w)=v_d$ shown in Figure~\ref{fig:delay-implicit-transform-just-v}.
	This approximation strongly suggests that $\pDer{w} g > 0$ in a neighborhood of the limit cycle, which we henceforth assume without proof.
	This assumption implies that the map $\paren{v,w}\mapsto \paren{v,v_d}$ is a diffeomorphism from a (possibly smaller) neighborhood of the limit cycle onto its image, and so the delay coordinates $\paren{v,v_d}$ yield a valid change of coordinates on this neighborhood.
		
	As in \S \ref{sec:standard}--\ref{sec:coord-inv}, let $\varphi\colon B\to S^1\subset \C$ be the asymptotic phase map defined on the basin of attraction $B$ of the limit cycle.
	The restriction of $\frac{1}{i\varphi}\pDer{v}\varphi|_{w=\textnormal{ const.}}$ to the limit cycle is the PRC curve associated with perturbing the system's voltage when the (unobserved) current remains fixed.
	If we were to mistakenly use $\frac{1}{i\varphi}\pDer{v}\varphi|_{v_d=\textnormal{ const.}}$ instead, our result will be in error by a term obtainable from the Jacobian of the coordinate transformation \eqref{eq:fhn-delay-coord-change}, as explained in Example~\ref{ex:delay-1} and demonstrated in the example of \S \ref{sec:ex-hopf}.
	
	We now describe a simulated experiment using this FHN model.
	First, we recorded a collection of steady-state oscillations which was assumed sufficiently large so as to enable a good estimate of the period of the limit cycle to our desired accuracy, and to enable modeling the limit cycle of the system in delay coordinates, as per \ref{sec:compute-gamma-y}, to our desired accuracy.
	We then allowed the system's state to evolve until we were confident it was close to the limit cycle, at which point we applied a small perturbing voltage, instantly driving the voltage slightly above or below the voltage when resting on the limit cycle while the current is unchanged.\footnote{
	    We note that many characterisations of neural oscillators in terms of phase do not apply a small voltage in order to estimate the PRC, but rather apply a spike (large voltage, short duration) to obtain a PRC which might be inconsistent with a first order approximation.
	    Such an experiment is not considered here, although we note that what is required to calculate the induced phase change across the stability basin is some method of integrating the accumulated phase change resulting from travelling from the limit cycle to the perturbed state.
	    Because the ensuing state change is typically ``large'', i.e., of the order of the size of the limit cycle itself, infinitesimal approximations are unlikely to be accurate.
	}
    We allowed the system to settle for a fixed amount of time equal to five periods (as estimated above) of the limit cycle, and then recorded the resulting phase shift.
    Because we waited an integer number of periods, the unperturbed base-point on the limit cycle would have returned to itself, simplifying the calculation.
		
	Using the data from this simulated experiment, we employed several ways of calculating the voltage PRC from this ensemble of trajectories. 
	First, we determined phase on the limit cycle as follows. 
	We took our estimate of the limit cycle and assigned to points on this curve a phase which advances uniformly in time using the algorithm \textit{Phaser} \cite{RevGuk08}.
	
	With phase on the limit cycle now a (numerically) known quantity, we assigned an approximate ``forward integration'' phase to a state by using the fact that we carefully estimated the period.
	The period was estimated by forward integration from a point on the limit cycle.
	The function returning the difference between the start and end point of such an integration was minimised using scipy's fmin function (with $10^{-12}$ tolerance for both the functions value and the location of the optima).
	The forward integration used the scipy function odeint with the Jacobian specified and low tolerances for the error control parameters atol and rtol, $10^{-1}$.
	
	For every trajectory, which by design all ended sufficiently close to the limit cycle, we estimated the phase of the final trajectory point from the phase of the nearest limit cycle points.
	To do this, the two closest points from the high resolution model of the limit cycle described above were selected and their phases noted.
	A line between these two points was constructed and the point whose phase was to be estimated was projected onto this line.
	The phase assigned to this point was that determined by a linear model of phase between the start and end point of the line.
	
	We then used the period and this phase to assign a phase to all other points on the trajectory, including the initial point.
	Since our perturbations were an instantaneous change in voltage from a point on the limit cycle, we took the phase of the perturbed initial point, subtracted from it the phase of the antecedent initial limit cycle point, and divided by the magnitude of the applied voltage. 
	Because the perturbation only changed $v$, but not $w$, this procedure gave us an estimate of the voltage PRC with current constant.
	
\begin{figure}[ht]
    \includegraphics[width=0.60\textwidth]{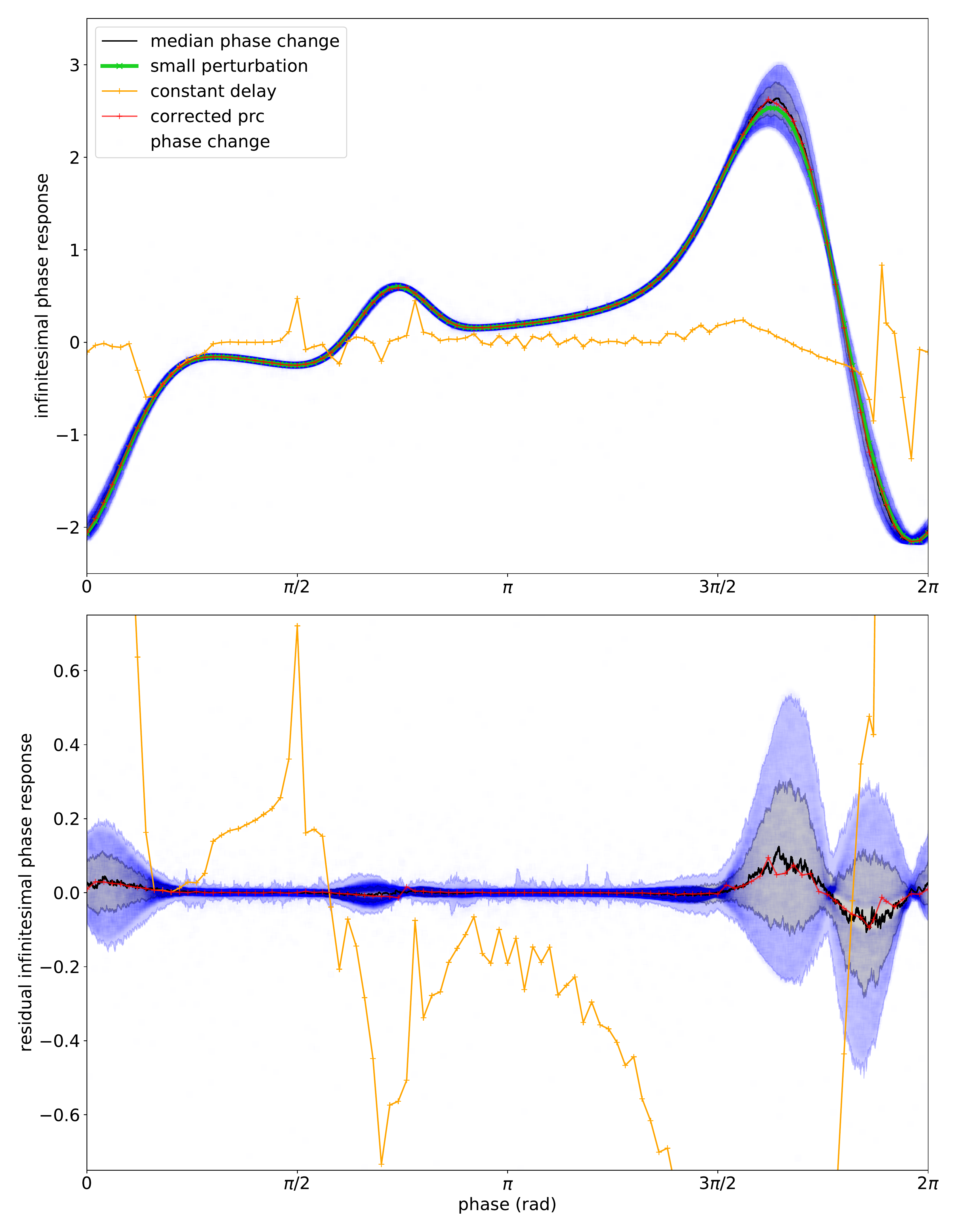}
    \caption{Plot of the normalised voltage PRC of the Fitzhugh-Nagumo system against the phase of the oscillator.
    We show a ground truth PRC (green) which we calculated numerically from a small perturbation, our sampled observations, a rolling median of the observations (black), $\frac{1}{i\varphi}\pDer{v}\phi|_{v_d=\textnormal{const.}}$ (orange), and $\frac{1}{i\varphi}\pDer{v}\phi|_{v_d=\textnormal{const.}}$ corrected with the Jacobian of the delay map to give $\frac{1}{i\varphi}\pDer{v}\phi|_{w=\textnormal{ const.}}$ (red).
    We computed the sampled observations by taking the difference between the phase a fixed time after the perturbation and the phase predicted by simulating the unperturbed system forward in time the same amount, dividing by the magnitude of the voltage perturbation, leading to variable results (blue dots; percentiles 2 and 98 light blue lines; 25 and 75 grey lines).
    Because some of the plots are hard to distinguish, we also plotted the same, but with the forward simulation (green, top subplot) subtracted (bottom subplot; all other line types unchanged).}
    \label{fig:calculatedfhnphase}
\end{figure}

	We followed this procedure twice.
	Once, on a deterministic FHN oscillator without any system noise, to give us a ``ground truth'' PRC (Figure \ref{fig:calculatedfhnphase}  ``small perturbation'' line).
	A second time, on simulated experimental data which includes system noise (Figure \ref{fig:calculatedfhnphase} ``phase change'' points, which show up a coloured region).
	Since the latter results were quantized by the nearest limit cycle end-point, and noisy, we also computed a windowed median of these point-wise estimates (Figure \ref{fig:calculatedfhnphase} ``median phase change'' line). 
	We used the median rather than mean since our system produced results with variability that seemed to be heavy tailed; rank statistics are more stable under such conditions than moments tend to be.

	\begin{figure}[ht]
    \includegraphics[width=0.9\textwidth]{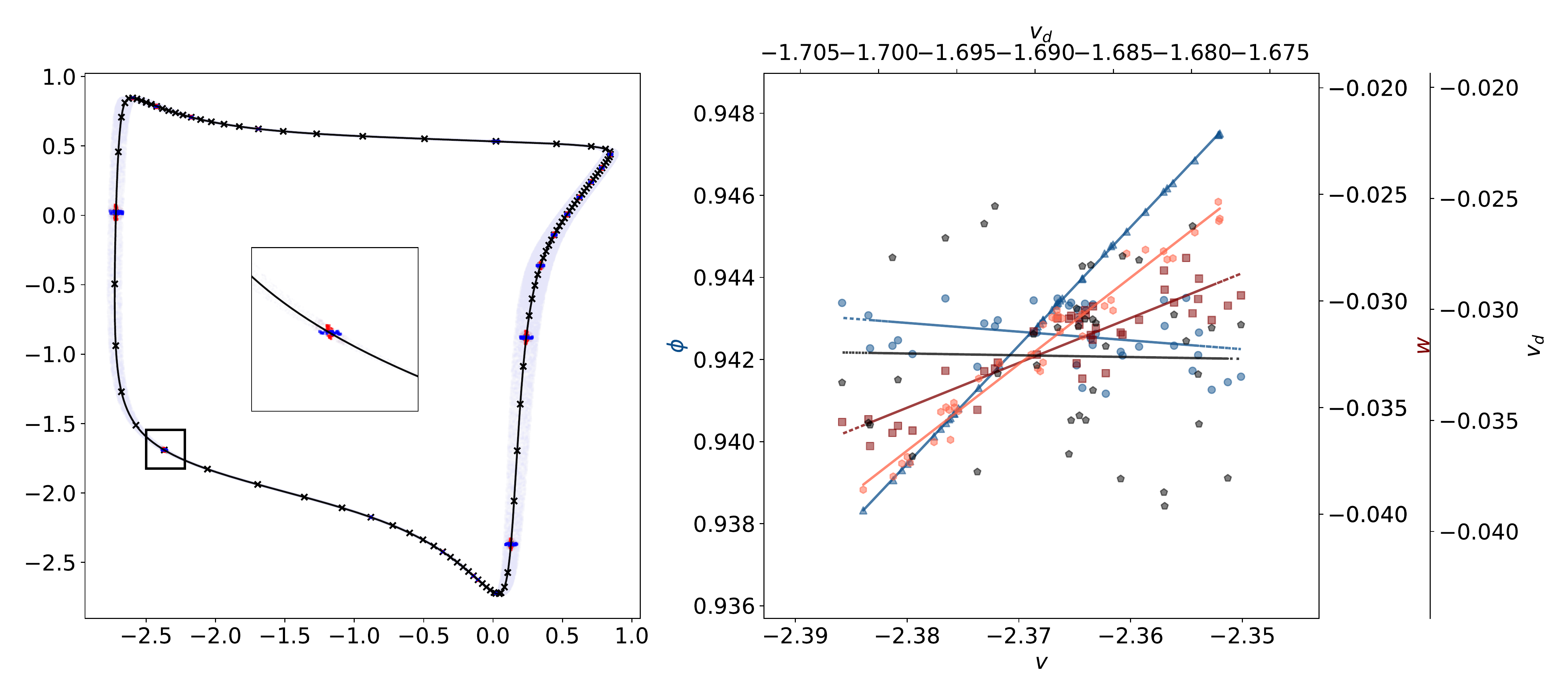}
    \caption{ %
    Process for calculating the PRC in delay coordinates by correcting using the Jacobian of the transformation from conventional to delay coordinates.
    We plotted the delay coordinate data for 100000 points (left subplot; sampled points in pale blue; x-axis is $v$, y-axis is $v_d$), indicated the limit cycle (black line) and the points at which we calculated the PRC (black crosses). %
    At every fifth such point we indicated vertical and horizontal subsets of points we used to compute the Jacobian with (gem blue and red). %
    We plotted an illustrative zoomed in view of one such location (boxed on limit cycle; center inset). %
    To demonstrate the quality of the various linear fits this process required, we plotted the individual linear input-output relationships (right subplot). %
    This includes the linear fits of the phase and the Jacobian of the transformation between $(v,v_d)$ and $(v,w)$ coordinates at the location of the inset. %
    The plot shows both functions of $v$ (with respect to the bottom x-axis; dotted lines), and functions of $v_d$ (top x-axis; solid lines). %
    We show phase change (blue; circular and triangular markers); $w$ change (orange and red; square and hexagonal markers); and $v_d$ (black; pentagonal markers).
    }
    \label{fig:calculationprocess}
    \end{figure}	
	
	\begin{figure}[ht]
    \includegraphics[width=0.60\textwidth]{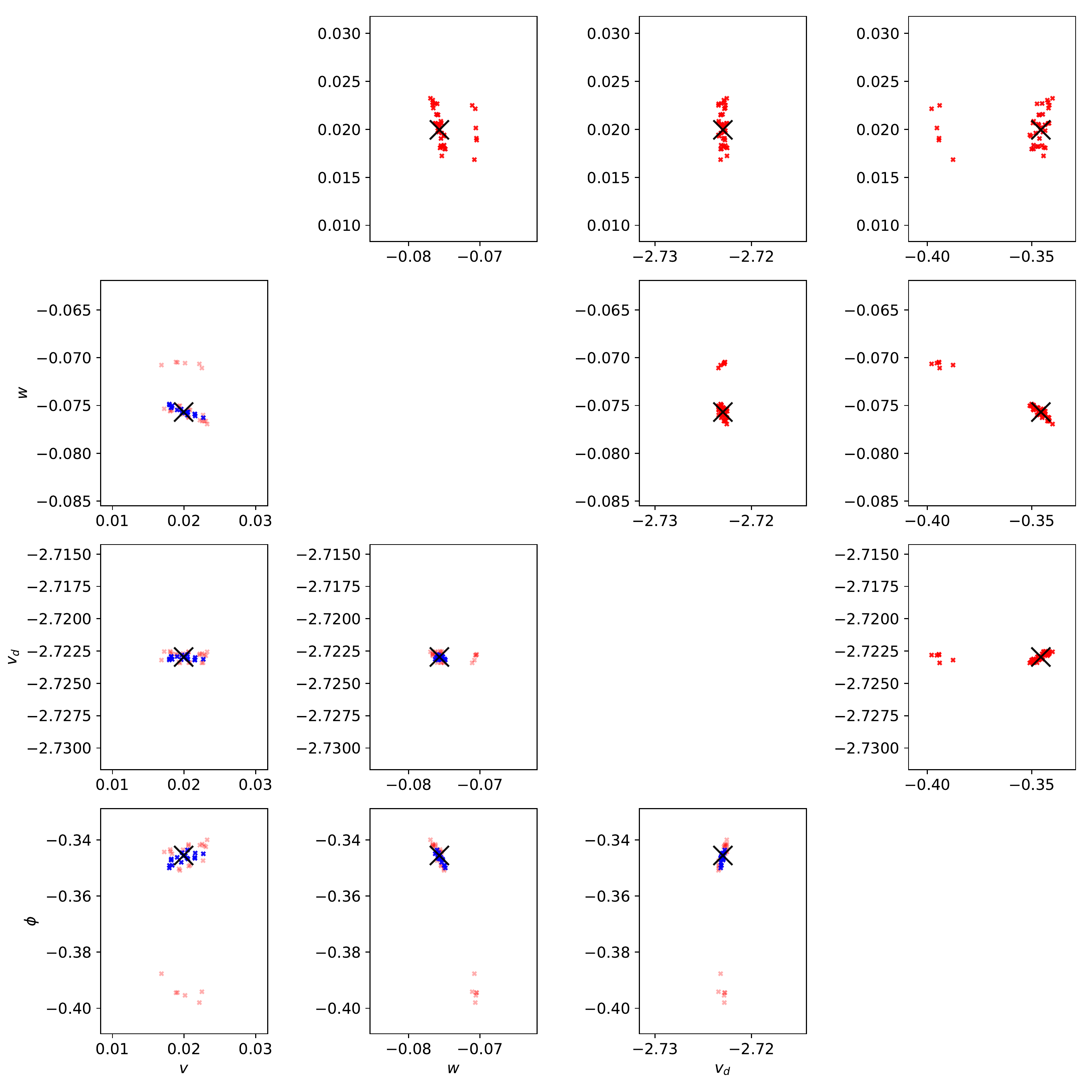}
    \caption{Scatter plot matrices of the dimensionless voltage, current, delayed voltage; and phase (red,blue) around a point (black) on the limit cycle. The transformation to delay coordinatees has a singularity in it which causes bimodality (red, top-right half of matrix, low alpha in the bottom-left); this is removed by excluding those observations which are seperated from the point about which we are calculating our derivatives (blue, bottom left). }
    \label{fig:combineddiagnostic}
    \end{figure}

For actual experimental data we should presume the presence of many types of noise: numerical, experimental system noise, and experimental measurement noise.
Under such conditions, the finite difference approximation of the derivatives we wished to compute would have poor statistical properties. 
A more noise-resilient method is to take small segments of our data at points around the limit cycle in delay coordinates, confine ourselves to slices of the data in the delay coordinate $v_d$, and perform a linear regression of the final phase against change in voltage $v$.
However, this method will calculate the PRC with a fixed delay coordinate (Figure \ref{fig:calculatedfhnphase} ``constant delay'' plot) which, for the reasons we explained in Example~\ref{ex:delay-1} and demonstrated in \S \ref{sec:ex-hopf}, will not be the voltage PRC with fixed current which we desire.
	
Recall that here we attempted to analyze numerically generated data as if this data came from experimental measurements, for which we have no underlying equations of motion \eqref{eq:fhn}.
Unlike in the example of \S \ref{sec:ex-hopf}, there is no way to  construct the delay coordinate transformation in closed form.\footnote{
    Even if we were assuming knowledge of the equations of motion, it seems likely to us that \eqref{eq:fhn} is sufficiently complicated that constructing the delay coordinate transformation in closed form is intractable. 
    We were able to construct the delay coordinate transformation in closed form for the toy example of \S \ref{sec:ex-hopf} since its form was chosen for this very purpose.	
}

Instead, we constructed a Taylor approximation of the delay coordinate transformation from the points in small regions.
Each region was centered at a limit cycle base-point and contained the $0.5\%$ observations in our training set nearest (by Euclidean distance) to that base-point (Figure \ref{fig:calculationprocess}).
Initially, we evaluated the Jacobian of the delay coordinate transformation by total least squares regression of the region points against their transformed images.
However, we discovered that this produced several numerical anomalies that we wished to explore further.

To make it easier for us to interpret and sanity-check the coordinate transformation, we further sub-divided the regions into high aspect-ratio horizontal and vertical strips (Figure \ref{fig:calculationprocess}, left sub-plot, blue and red).
For a region sufficiently small to have a nearly constant Jacobian, a scatter plot of the major axis coordinate of a strip against a transformed output coordinate value is expected to be a line segment whose slope is the respective element of the Jacobian (Figure \ref{fig:combineddiagnostic} bottom $4 \times 4$ grid).
Inspection of this kind of scatter plot for other than linear structure allowed us to resolve the numerical issues we encountered.

\FloatBarrier

In most cases, we used the following procedure.
As initialized, each of the strips contained at most $20\%$ of the points in the regions, but this fraction was adjusted later.
Using linear regression, we calculated the PRC for both $v$ and $v_d$, with $v_d$ and $v$ held constant, respectively. 
In addition, we calculated the Jacobian of the transformation from $(v,w)$ to $(v,v_d)$ by linear regression. 
We then reduced the size of the region of admitted points until the $R^2$ statistic of the linear fit was increased above an arbitrarily chosen threshold of $0.75$.
This automatic scaling ensured that we were sampling a region within which the higher order Taylor polynomial terms do not dominate the results.

However, in some regions the coordinate transformation for the delay we chose had singularities (``folds'') coming close to the limit cycle.
As a consequence, the image points of some of the strips were clearly quadratic (see Figure \ref{fig:dependentquadraticexample} or bi-modal (Figure \ref{fig:combineddiagnostic} top $4 \times 4$ grid).
In these cases too the auto-scaling helped prevent the numerical problems we initially encountered.

Armed with a high quality estimate of the Jacobian of the coordinate change $(v,w) \mapsto (v,v_d)$, we calculated the desired PRC $\frac{1}{i\varphi}\pDer{v}\varphi|_{w=\textnormal{const.}}$ using the $(v,v_d)$ coordinates and the estimated Jacobian as in Example~\ref{ex:delay-1} and in \S \ref{sec:ex-hopf} (see Figure \ref{fig:calculatedfhnphase} ``corrected prc'').
There is excellent agreement between the results of this calculation, using simulated experimental data with delay coordinates, and that obtained numerically by the other calculation methods.
Furthermore, the computation we performed allowed us to obtain a voltage PRC with current constant that very closely follows that which we could have obtained by computing the median of an enormous number of experiments (see Figure \ref{fig:calculatedfhnphase} bottom sub-figure for detail of differences between PRC calculation results).
It is crucial to note that this entire simulated experimental protocol did not require any measurements of current.

\section{Conclusion}
We have pointed out both the danger and opportunity inherent in attempting to construct oscillator models including their phase response curves from measuring only a subset of the state variables.
Such an approach could allow us, for example, to fully characterize an FHN-like neuron by only measuring and perturbing ``voltage''.
However, the resulting PRC will then be correct only with respect to the delay embedding we use to describe the dynamics.
Since in practice, some other non-observed quantities (such as ``current'') are held constant while coupling to the neuron, we then showed how a model of those quantities can be used to convert the PRC to the correct coordinates.
We demonstrated this computation in a form that is robust to noise and potentially suitable for the analysis of experimental data.
This process can be fraught with numerical difficulties if the coordinate change required is singular close to the limit cycle, and we have shown techniques which have helped us ascertain when and where such problems appeared in our data.

Our approach provides a rapid method for obtaining a PRC in cases where the unobserved dynamics (current in our neuroscience example) are well known.
However, from a neuroscience perspective there is an important lesson to be learned here: for any given set of voltage measurements, there can be a variety of current models which could be assumed with equal validity, and the resulting PRC would accordingly change.
Our approach gives the ``correct'' PRC for an experimental system if the model of the state variable that remained constant under perturbation is correct.
If the model is not correct, there is no guarantee that the PRC will be accurate.
This dependency might be a means for eliminating among multiple current generation models, provided the voltage measurements provide an alternative method for measuring the PRC against which the predictions from different current models can be tested.
As new current generation dynamics are hypothesized, it may be possible to use the differences in PRC that would be produced by these new models to discern which ``current'' is consistent with a given neuron's pre-recorded voltage measurements --- all without requiring any new data collection. 

The approach we present here is by no means limited to neuroscience applications. 
Future applications of this work include producing reliable and robust models of coupled oscillator systems in any domain in which oscillator models are used.

In this paper we considered only classical asymptotic phase for deterministic oscillators.
However, there are also multiple notions of stochastic phase defined for stochastic oscillatory systems \cite{schwabedal2013phase,thomas2014asymptotic,cao2020partial,engel2021rdsphase} (see \cite{pikovsky2015comment,thomas2015reply} for a spirited discussion), and it would be interesting to investigate these notions in the context of the present work.

\subsection*{Acknowledgments}
Kvalheim was supported in part by the Army Research Office (ARO) under the SLICE Multidisciplinary University Research Initiatives (MURI) Program, award W911NF-18-1-0327, and in part by ONR N000141612817, a Vannevar Bush Faculty Fellowship held by Daniel E. Koditschek, sponsored by the Basic Research Office of the Assistant Secretary of Defense for Research and Engineering.
Revzen was supported by ARO W911NF-14-1-0573, ARO MURI W911NF-17-1-0306, NSF CMMI 1825918, and the D. Dan and Betty Kahn Michigan-Israel Partnership for Research and Education.
	
	\bibliographystyle{amsalpha}
	\bibliography{all}
	
\begin{appendix}
\section{Estimating the period of the Fitzhugh-Nagumo system}
Below is a listing of the python code used to calculate the period of the Fitzhugh-Nagumo oscillator. This oscillator is slow-fast and quite stiff, so some care is needed when integrating it; the default settings of the scipy odeint command do not produce reliable results.

It is useful to supply the Jacobian and to have a large number of intermediate steps. In addition, for the precision we required for the period in this work, a higher than default tolerance was needed.

\begin{minted}{python}
from numpy import stack,linspace
from scipy.integrate import odeint
from scipy.optimize import fmin

# Derivatives and partial derivatives of the Fitzhugh-Nagumo system.
def vdot(v,w,z,c):
  return(c*(w+v-((v*v*v)/3.)+z))

def wdot(v,w,a,b,c):
  return(-(v-a+b*w)/c)

def vdotdv(v,w,z,c):
  return(c*(1.-v*v))

def vdotdw(v,w,z,c):
  return(c)

def wdotdv(v,w,a,b,c):
  return(-1./c)

def wdotdw(v,w,a,b,c):
  return(-b/c)

def xdotfhn(p,t=0,z=-0.4,a=0.7,b=0.8,c=3.,tau=1.):
  '''
  Vectorised form of the Fitzhugh-Nagumo system
  '''
  x,y = p[...,0],p[...,1]
  return(tau*stack([
    vdot(x,y,z,c),
    wdot(x,y,a,b,c)
  ],-1))

def Jxfhn(p,t=0,z=-0.4,a=0.7,b=0.8,c=3.,tau=1.):
  '''
  Jacobian of the Fitzhugh-Nagumo system (vectorised)
  '''
  x,y = p[...,0],p[...,1]
  return(tau*stack([
    stack([vdotdv(x,y,z,c),vdotdw(x,y,z,c)],-1),
    stack([wdotdv(x,y,a,b,c),wdotdw(x,y,a,b,c)],-1)
  ],-1))

def loopf(tc):
  '''
  Integrate the Fitzhugh-Nagumo system for a period of tc and return the 
  difference between the initial and final point. 
  Minimising this will give an integral multiple of the period.
  '''
  y1 = odeint(xdotfhn,y0,[0,tc],Dfun=Jxfhn,col_deriv=True,rtol=tol, atol=tol, mxstep=mxstep)[-1]
  return(sum((y0-y1)**2))


tc0 = 3*3.743     # Initial guess at the period
tol = 1.e-10      # Tolerance
trelax = 1000     # A long period of time, sufficient for the system to relax
Nstep = 1000      # Need extra steps for long initial integration
mxstep=5000       # Maximum number of intermediate steps for integrator
yinit = [1.0,0.0] # A starting location in stability basin

# Get a point on the limit cycle by integrating for a long period of time
y0 = odeint(
  xdotfhn,
  yinit,
  linspace(0,trelax,Nstep),
  Dfun=Jxfhn,
  col_deriv=True,
  rtol=tol,
  atol=tol,
  mxstep=mxstep
)[-1]

# Minimise the squared difference between the starting position on the limit 
# cycle and the final position on the limit cycle by adjusting the integration 
# period, should have a minimum of zero to within tolerace after convergence.
tc1 = fmin(loopf,tc0,ftol=1e-12,xtol=1e-12)[0]
print("Period estimated to be:",tc1)
\end{minted}

\section{Special cases estimating the infinitesimal PRC}
There are multiple special cases which must be considered when estimating the partial derivatives used to calculate the infinitesimal phase response curve close to the limit cycle for this system.

We consider two models. 
First, we select a model where the dependent variable (the phase for the infinitesimal phase response curve, the un-transformed coordinates for the Jacobian) is polynomial of the independent variable (the transformed coordinate).
We consider the linear and quadratic cases, and select from these models via an AIC.
An example where the linear model was superior is included in supplementary figure \ref{fig:linearexample}.

\begin{figure}[ht]
\includegraphics[width=0.90\textwidth]{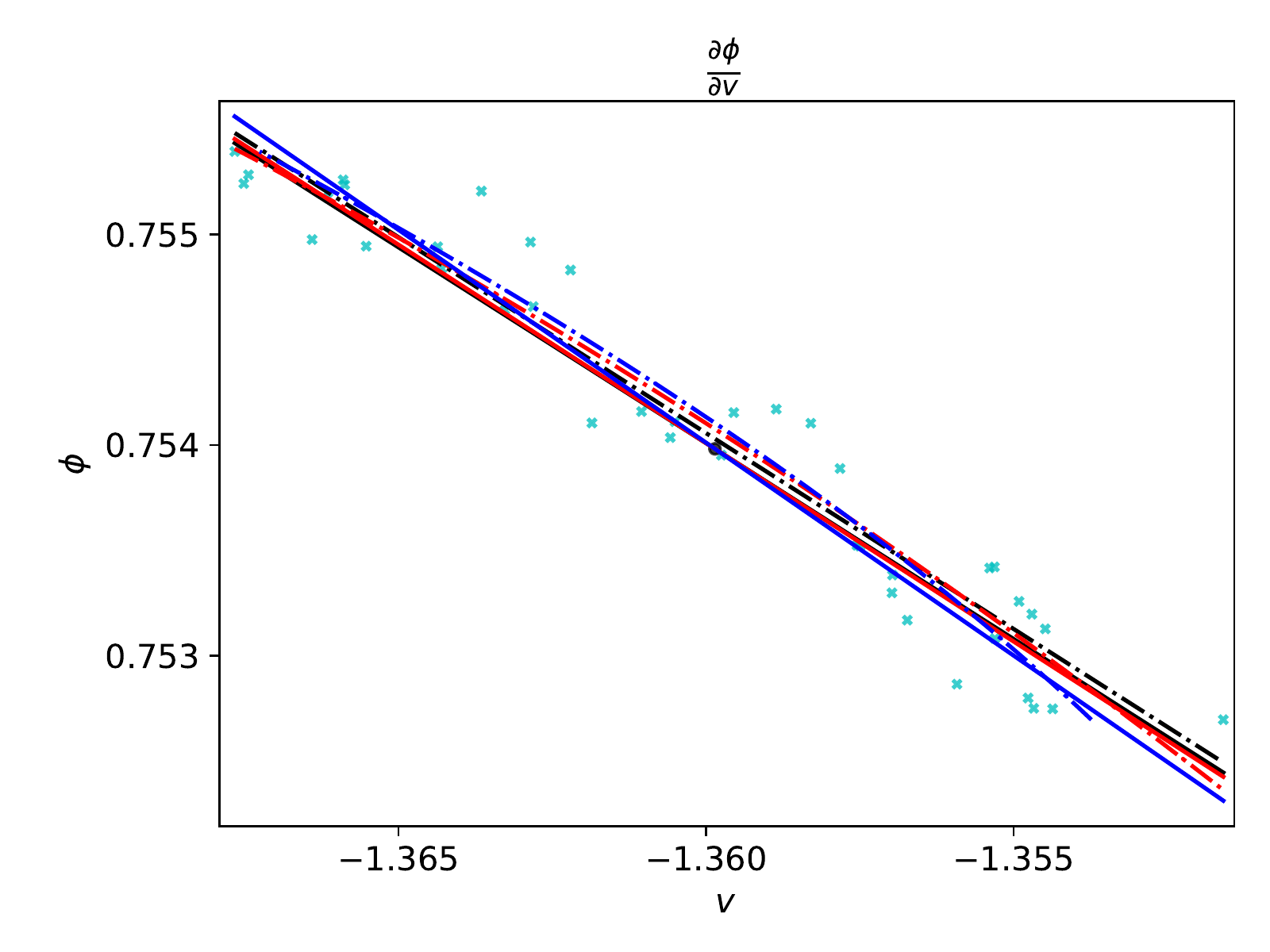}
\caption{%
An example where a simple linear model was used to estimate the rate of change of the phase against change in dimensionless voltage at a fixed point on the limit cycle (black point in center of plot). %
We plotted the phases of the points at fixed delay voltages $v_d$ across a range of voltages $v$ (cyan crosses). %
We also plotted various model fits (dot-dash lines): linear (black), quadratic in $v$ (red), and quadratic in phase $\phi$ (blue). %
These imply potentially unequal derivatives which we indicate as lines with the associated slope (same color, solid thin lines). %
In this instance all fits give comparable estimates for the slope.}
\label{fig:linearexample}
\end{figure}

For some cases this model is deficient. There is a clear non-linear relationship between the independent variable and dependent variable, such that a quadratic in the dependent variable is a more appropriate model. 
Such cases can be readily identified by inspection, the non-linearity is obvious and there is frequently bi-modality in the distribution of the dependent variable. 
In such cases a quadratic in the dependent variable is instead used.
This is illustrated in supplementary figure \ref{fig:dependentquadraticexample}.

\begin{figure}[ht]
\includegraphics[width=0.90\textwidth]{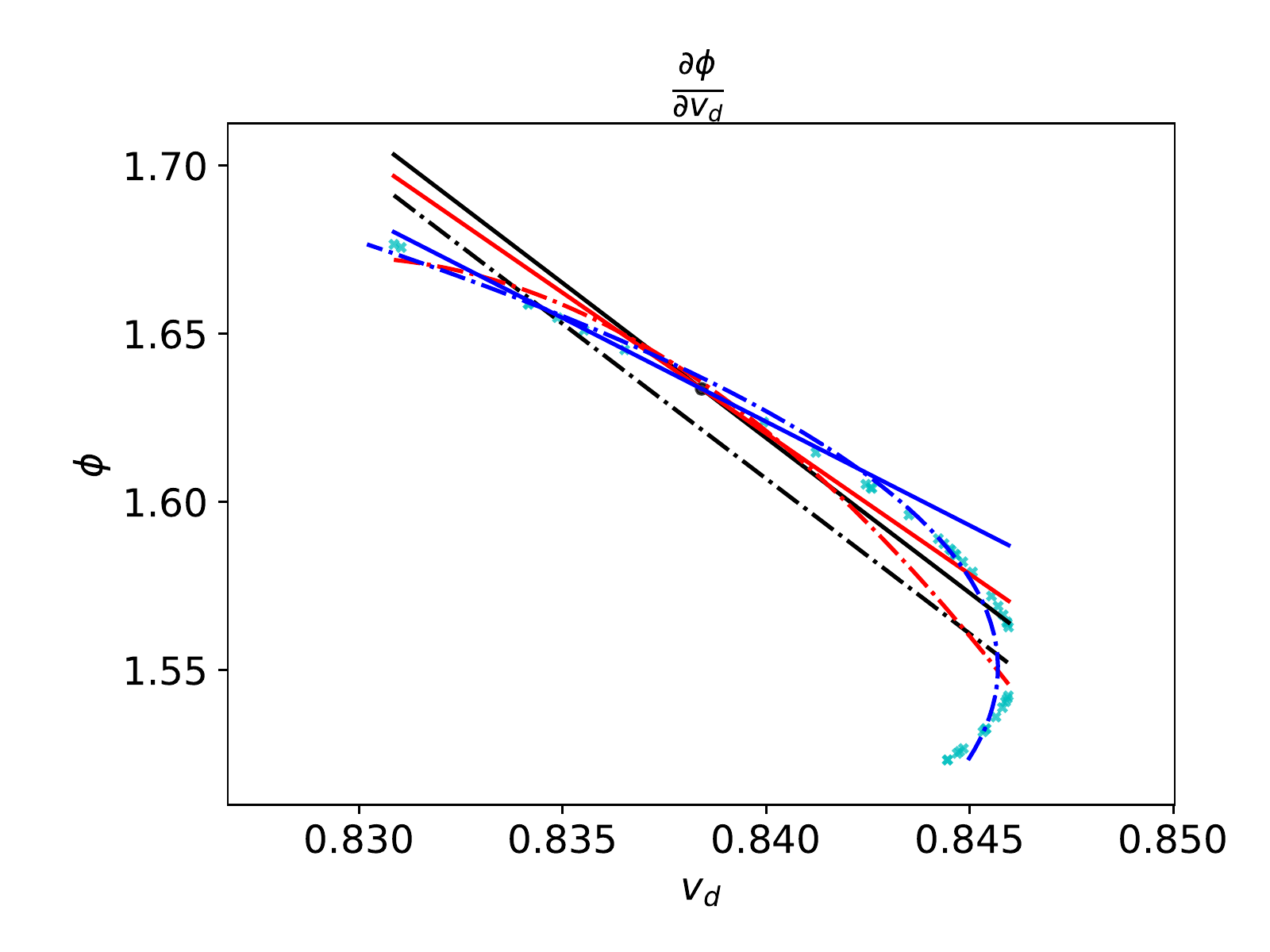}
\caption{An example where there is quadratic dependence of phase against delayed dimensionless voltage. The elements of this plot are as in Figure \ref{fig:linearexample}. %
The model has this quadratic dependence is clearly superior. }
\label{fig:dependentquadraticexample}
\end{figure}

In some cases the transformation from the conventional coordinates of the Fitzhugh-Nagumo system to the delay coordinates contains a singularity which is very close to the limit cycle.
This was handled in two ways.

In most cases the approach of this singularity was close, but not so close that a reasonable estimate of the gradient could not be obtained by simply excluding manually those points which came excessively close to the limit cycle.
An example of such a case is illustrated in supplementary \ref{fig:singularityexample}.

\begin{figure}[ht]
\includegraphics[width=0.90\textwidth]{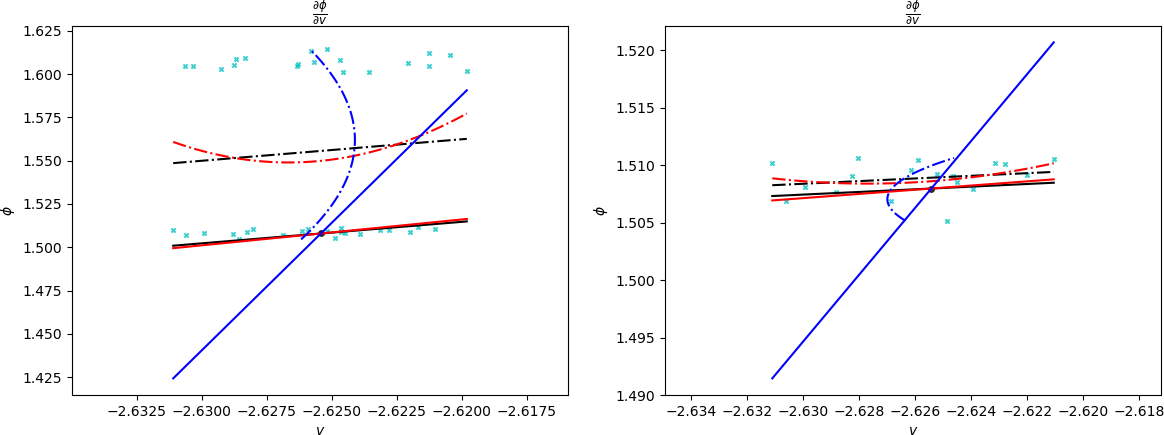}
\caption{ %
  An example where a singularity in the transformation to delay coordinates is present and close to the limit cycle. %
   The elements of this plot are as in Figure \ref{fig:linearexample}. %
  In this case, those points which are on the wrong side of the singularity can simply be excluded (right panel) and a better estimate of the partial derivative is obtained.
 }
\label{fig:singularityexample}
\end{figure}

For one test point the singularity came extremely close to the limit cycle and no reasonable estimate of the phase response was possible. Instead two additional points were selected near to this example and the phase response estimated at these locations instead.
This is illustrated in supplementary figure \ref{fig:element94}.

\begin{figure}[ht]
\includegraphics[width=0.90\textwidth]{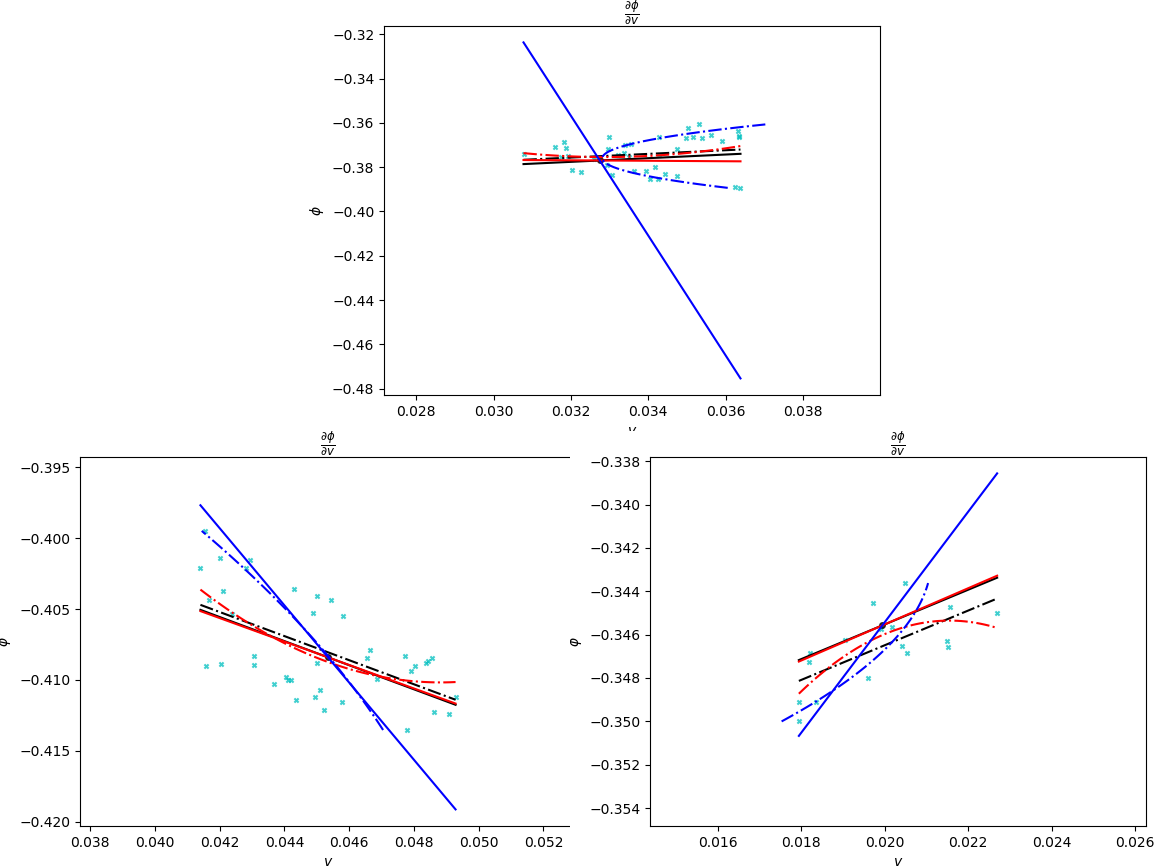}
\caption{Here, the singularity in the transformation to delay coordinates is too close to the limit cycle, and no reasonable estimate of the partial derivative can be obtained, illustrated in the top panel.  %
The elements of this plot are as in Figure \ref{fig:linearexample}. %
Instead, two points (bottom left and right panels) close by on the limit cycle are considered and estimates for the partial derivatives are obtained there. }
\label{fig:element94}
\end{figure}
\end{appendix}
	
\end{document}